\documentclass[aps,twocolumn,showpacs,prl,floatfix,superscriptaddress,longbibliography]{revtex4-1}
\usepackage{amsmath}
\usepackage{amsthm}
\usepackage{amsfonts}
\usepackage{bbm}
\usepackage{color}
\usepackage{graphicx}
\usepackage{dcolumn}
\usepackage{array}
\usepackage{float}
\usepackage{supertabular}
\usepackage{longtable}
\usepackage{txfonts}
\usepackage{wasysym}
\usepackage{cases}
\usepackage[T1]{fontenc}
\usepackage[latin1]{inputenc}
\usepackage{amssymb}
\usepackage[usenames,dvipsnames]{xcolor}
\usepackage{amstext}
\usepackage{latexsym}
\usepackage[colorlinks=true,citecolor=Cerulean,linkcolor=RubineRed,urlcolor=Cerulean]{hyperref}
\usepackage{enumitem}
\usepackage{epsfig}
\usepackage{dsfont}
\usepackage{mathrsfs}
\usepackage{arydshln,leftidx,mathtools}
\usepackage{color}
\usepackage{amsmath}

\DeclareMathOperator{\sign}{sgn}

\begin{document}
\title{Exactly-solvable system of  one-dimensional trapped bosons with short and long-range interactions}
\author{M. Beau}
\affiliation{Department of Physics, University of Massachusetts, Boston, Massachusetts 02125, USA}
\affiliation{Dublin Institute for Advanced Studies, School of Theoretical Physics, 10 Burlington Road, Dublin 4, Ireland}
\author{S. M. Pittman}
\affiliation{Department of Physics, Harvard University, Cambridge, MA 02138, USA}
\author{G. E. Astrakharchik}
\affiliation{Departament de F\'isica, Universitat Polit\`ecnica de Catalunya, Campus Nord B4-B5, E-08034, Barcelona, Spain}

\author{A. del Campo}
\affiliation{Donostia International Physics Center, E-20018 San Sebasti\'an, Spain}
\affiliation{IKERBASQUE, Basque Foundation for Science, E-48013 Bilbao, Spain}
\affiliation{Department of Physics, University of Massachusetts, Boston, MA 02125, USA}
\affiliation{Theory Division, Los Alamos National Laboratory, MS-B213, Los Alamos, NM 87545, USA}

\def\i{{\rm i}}
\def\L{{\rm \hat{L}}}
\def\q{{\bf q}}
\def\l{\left}
\def\r{\right}
\def\te{\mbox{e}}
\def\d{{\rm d}}
\def\t{{\rm t}}
\def\K{{\rm K}}
\def\N{{\rm N}}
\def\H{{\rm H}}
\def\la{\langle}
\def\ra{\rangle}
\def\om{\omega}
\def\Om{\Omega}
\def\vep{\varepsilon}
\def\wh{\widehat}
\def\tr{{\rm Tr}}
\def\da{\dagger}
\def\iz{\left}
\def\zi{\right}
\newcommand{\beq}{\begin{equation}}
\newcommand{\eeq}{\end{equation}}
\newcommand{\beqa}{\begin{eqnarray}}
\newcommand{\eeqa}{\end{eqnarray}}
\newcommand{\intf}{\int_{-\infty}^\infty}
\newcommand{\into}{\int_0^\infty}

\begin{abstract}
We consider  trapped bosons with contact interactions as well as Coulomb repulsion or gravitational attraction in one spatial dimension. The exact ground state energy and wave function are identified in closed form together with a rich phase diagram, unveiled by Monte Carlo methods, with crossovers between different regimes. A trapped McGuire quantum soliton describes the attractive case. Weak repulsion results in an incompressible Laughlin-like fluid with flat density, well reproduced by a Gross-Pitaevskii equation with long-range interactions. Higher repulsion induces Friedel oscillation and the eventual formation of a Wigner crystal. 

\end{abstract}

\maketitle

Low-dimensional quantum gases can be engineered with ultracold atomic vapors by freezing the dynamics along a given axis under tight confinement.
In one spatial dimension,  quantum fluctuations are enhanced, and yet, the stability of the system extends   to strongly-interacting regimes, in contrast with the three-dimensional case. Effectively one-dimensional quantum gases offer a test-bed for integrability and provide a faithful implementation of a range of exactly solvable models \cite{Cazalilla11,Guan13}. Among them, the homogeneous gas with contact interactions described by a $\delta$-function pseudopotential, exactly solved in 1963 by Lieb and Liniger (LL model)~\cite{LL63,L63}, is a paradigmatic reference model.  Olshanii showed that ultracold atomic gases confined in tight waveguides experience resonance in the effective one-dimensional interactions and can be used to realize the LL model \cite{Olshanii98,LSY03}. Following this observation, the LL model has been routinely  realized in the laboratory \cite{Tolra2004, Kinoshita2004, Kinoshita2005, Haller2009, Armijo2010, Haller2011, Jacqmin2011, Fabbri2015, Meinert2015, Atas2017, Wilson2020}.

From the theoretical point of view, a rich variety of powerful mathematical methods such as the Bethe ansatz is available to analyze LL model, making it a favorite test-bed for these techniques \cite{KBI97,Takahashi99}.
The limit of strong interactions, known as the Tonks-Girardeau gas, corresponds to hard-core bosons~\cite{Girardeau60} and has helped to elucidate the fact that quantum exchange statistics is ill-defined as an independent concept in one dimension, being  inextricably woven to inter-particle interactions. Its study revealed the existence of Bose-Fermi duality and its generalizations~\cite{Girardeau06,Batchelor06}.
More recently, it has been pointed out that the LL gas constitutes the universal nonrelativistic limit of a variety of integrable quantum field theories~\cite{Mussardo16}.
In nonlinear physics, the quantum version of both bright and dark solitons has been found in the LL model. In particular, the cluster solution of the attractive LL gas was explicitly found by McGuire via Bethe ansatz~\cite{McGuire64}. For repulsive interactions, many-body states describing gray solitons, characterized by a dip in the density profile of finite depth, have been discussed in  \cite{Kulish76,Ishikawa80,Deguchi1,Deguchi2,AstrakharchikPitaevskii2013,Kaminishi19}. Dark solitons with vanishing density at the dip have been also found in the Tonks-Girardeau regime that describes hard-core bosons~\cite{GW00,AstrakharchikPitaevskii2013,Reichert2019}.

The LL model is known to admit an exact treatment by Bethe ansatz in homogeneous external potentials such as a ring~\cite{LL63}, a box~\cite{Gaudin71,Batchelor05,PhysRevLett.123.250602}, or the continuum~\cite{Gaudin14}. While the Bethe ansatz method has proved to be very powerful and useful~\cite{KBI97,Takahashi99}, it cannot account exactly for the presence of a non-uniform confinement that is ubiquitous in experimental settings with trapped ultracold gases. In the presence of a harmonic trap, away from the Tonks-Girardeau limit~\cite{GWT01}, the understanding of the LL gas relies on a combination of approximate, effective theories and numerical methods~\cite{Cazalilla11,AG02,GAZ15}. In this context, it is thus of great importance to find other exactly-solvable many-body models in a trapped geometry.

In this work we introduce a novel exact solution for a one-dimensional Bose gas in the presence of a harmonic trap and with both short-range contact and long-range linear pairwise inter-particle interactions. The long-range contribution corresponds physically to one-dimensional Coulomb repulsion or  gravitational attraction. The exact ground state is found in closed form and is shown to reduce to the McGuire soliton solution for attractive interactions when the frequency of the harmonic confinement vanishes. We characterize the ground-state properties and demonstrate that a wide variety of regimes can be accessed in this system, including soliton-like, ideal Bose gas, incompressible fluid and Wigner-crystal regimes.

We start by considering a very general system consisting of $N$ atoms of mass $m$ confined in a harmonic trap of frequency $\om$, interacting via both the contact interaction with coupling strength $g$ and a long-range potential of strength $\sigma$. The system Hamiltonian is 
\beqa
\label{LRLL}
H=\sum_{i=1}^{N}\left( -\frac{\hbar^2}{2m}\frac{\partial^2}{\partial x_i^2} +\frac{1}{2}m\omega^2x_i^2\right)
+\!\sum\limits_{i<j}^N \left[g\delta(x_{ij})+\sigma|x_{ij}|\right]\!,
\eeqa
where $x_i$ are particle coordinates and $x_{ij}=x_i-x_j$ the relative distances. The long-range interaction potential in Eq.~(\ref{LRLL}) corresponds to the solution of the Poisson equation for a point source distribution of mass or charge $\Delta V({\bf r})=\kappa \delta({\bf r})$. For any dimension $d\geq 1$ the Fourier transform of the Laplacian equals to $-{\bf k}^2$, where ${\bf k}\in\mathbb{R}^d$ is the wave vector. For a point source distribution of mass or charge, the Fourier transform of the static potential $V$ is given by $\pm Q/{\bf k}^2$, where $Q$ is the electric charge or mass of the particles. Taking the inverse Fourier transform yields $V=\mp Q|x|$ in one-dimension (the familiar case $V=\pm Q/|{\bf r}|$ is found in $d=3$).

The exchange operator formalism introduced by Polychronakos~\cite{Polychronakos92} has brought great insight into one-dimensional quantum integrable models and we use it to analyze Hamiltonian~(\ref{LRLL}). We show that the latter admits a decoupled form, in terms of phase-space variables. To do this, we consider the Hermitian generalized momenta of $N$ particles 
\begin{eqnarray}\label{Gmomentum}
\pi_i=p_i + \i \sum_{j\neq i}V_{ij}M_{ij},
\end{eqnarray}
defined in terms of the canonical momenta $p_j=-\i\hbar\frac{\partial}{\partial x_j}$ for particles $ j=1,\cdots,N$. The particle permutation operator $M_{ij}$ is idempotent $M_{ij}^2=1$, symmetric $M_{ij}=M_{ji}$ and acts on an arbitrary operator $A_j$ as $M_{ij}A_j=A_iM_{ij}$, $M_{ij}A_k=A_k M_{ij}$.
We choose the so-called prepotential function $V_{ij}=V(x_{ij})$ to be proportional to the sign function  
\begin{equation}\label{prepot}
V_{ij}=-\frac{\hbar}{a_s}\sign{(x_{ij})},
\end{equation}
with $a_s$ being a real constant (negative or positive) with units of length. Later, we shall see that $a_s$ is physically equivalent to a $s$-wave scattering length.

First we consider a homogeneous case and construct a purely kinetic Hamiltonian of the form,
\begin{eqnarray}
H_0&=&\frac{1}{2m}\sum_{i}\pi_i^2\\
& =& \sum_i \frac{p_i^2}{2m} + \frac{1}{2m}\left(\sum_{i < j}(\hbar V'_{ij}M_{ij} + V^2_{ij})-2\sum_{i< j < k} V_{ijk}M_{ijk}\right),\nonumber
\end{eqnarray}
\noindent 
where $V_{ijk}=V_{ij}V_{jk}+V_{jk}V_{kl}+V_{kl}V_{ij}$ , $M_{ijk}=M_{ij}M_{jk}$, and the prime denotes the spatial derivative. Explicit computation shows that $V^2_{ij}=\hbar^2/a_s^2$ and $V'_{ij}=-(2\hbar/a_s)\delta(x_{ij})$ for the two-body term, while the three-body term reduces to a constant $V_{ijk}= -\hbar^2/a_s^2$. This yields the many-body Hamiltonian of a one-dimensional gas with pairwise contact interactions
\beqa
\label{HLL}
H_{\rm LL} = H_{0}+E_0 &=& -\frac{\hbar^2}{2m}\sum_{j=1}^{N} \frac{\partial^2}{\partial x_j^2} 
+ g\!\sum_{i < j}\delta(x_{ij})M_{ij},
\eeqa
where the coupling constant is related to the one-dimensional $s$-wave scattering length as $g = -2\hbar^2/(ma_s)$ and 
\beqa
E_0 
= -\frac{m g^2}{\hbar^2}\frac{N(N^2-1)}{24}. 
\eeqa 
For Bose statistics, $M_{ij}$ reduces to the identity and Hamiltonian~(\ref{HLL}) describes $N$ one-dimensional bosons subject to $s$-wave pairwise interactions, i.e., Lieb-Liniger~\cite{LL63,L63} and McGuire~\cite{McGuire64} systems. It is known that for repulsive interactions, $g>0$, only scattering states are possible and the homogeneous gas is stable. On the contrary, for attractive interactions, $g<0$, the system collapses into a many-body bound state   that describes a bright quantum soliton, and thus, the thermodynamic limit does not exist~\cite{McGuire64}. 

To include a harmonic trap, it is convenient to introduce analogues of the creation and annihilation operators
\begin{eqnarray}
a_{i}=\frac{\pi_i-\i m\omega x_i}{\sqrt{2m\omega\hbar}},\ a_{i}^{\dagger}=\frac{\pi_i+\i m\omega x_i}{\sqrt{2m\omega\hbar}},
\end{eqnarray}
satisfying
\begin{eqnarray}
[a_i,a_i^\dagger]
& = & 1 + \frac{m g}{\hbar^2}\sum_{j\neq i}|x_{ij}|M_{ij}.  
\end{eqnarray}
The relation $\frac{\hbar\omega}{2}\sum_{i} \{a_i,a^{\dagger}_i\} = H_{LL}+\sum_{i}\frac{1}{2} m\omega^2 x_i^2$ 
and the identity $a_i^\dagger a_i=\frac{1}{2}\{a_i,a_i^\dagger\}-\frac{1}{2}[a_i,a_i^\dagger]$ 
allows us to derive the Hamiltonian of the system embedded  in a trap,
\begin{eqnarray}\label{HLLhl}
H&=&\hbar \omega \sum_{i} a_i^\dagger a_i + E_0  \\
&=&\sum_{i=1}^{N}\left(-\frac{\hbar^2}{2m}\frac{\partial^2}{\partial x_i^2}
+\!\frac{m\omega^2x_i^2}{2}\right)
+\!\sum_{i<j}\left[g\delta(x_{ij})
\!-\!\frac{m\omega g|x_{ij}|}{\hbar}
\right]\;.\nonumber
\end{eqnarray}
This is an instance of the Hamiltonian class ~(\ref{LRLL}) in which the trap frequency $\omega$, the coupling constant $g$ and the strength of the long-range interaction $\sigma$ satisfy the following relation
\begin{equation}
\label{Eq:sigma}
\sigma = -m\omega g/\hbar.
\end{equation}
Thus, for repulsive contact interactions ($g>0$), the linear long-range term corresponds to  Coulomb repulsion between equal charges in one dimension. Similarly, for attractive contact interactions ($g<0$) the linear long-range term describes gravitational attraction between equal masses in $d=1$. Thus, short and long-range interactions are either both attractive or both repulsive. 

The value of the coupling $\alpha=\omega g/\hbar$ depends on the frequency of the trap, the mass of the particle and the coupling constant $g$. Using ultracold gases as a platform, one could adjust the value of $g$ independently of the mass of the particles using a Feshbach resonance. This is tantamount to tuning the effective gravity/anti-gravity acceleration $\alpha$ or electrostatic charge $\hbar\omega c/(4\pi \epsilon_0 r_\perp^2)$.

The exact expression for the ground-state energy of Hamiltonian~(\ref{HLLhl}) can be written explicitly as
\begin{equation}
E_0=\frac{N\hbar \omega}{2}-\frac{mg^2}{\hbar^2}\frac{N(N^2-1)}{24}\;,
\label{Eq:E}
\end{equation}
which is independent of the sign of the coupling constant $g$, and scales as $\propto N^3$ for large atom number as will be commented in more details later. For bosons ($M_{ij}=1$), the ground-state wave function satisfies $a_i \Psi_0 (\boldsymbol{x})=0$, which can be written in terms of the logarithmic derivative, 
\begin{eqnarray*}\label{CalcGS1} 
\frac{\partial_{x_i} \Psi_0}{\Psi_0}
=
-\frac{x_i}{a_{ho}^2}
- \sum_{j\neq i}\frac{\sign(x_{ij})}{a_s}\;,
\end{eqnarray*} 
\noindent
where $a_{ho}=\sqrt{\hbar/(m\omega)}$ is the harmonic oscillator length. Hence, it follows that the exact ground-state wave function of Hamiltonian~(\ref{LRLL}) is
\begin{multline}\label{Psi0HLLhl}
\Psi_0(\boldsymbol{x})=
\mathcal{N}^{-1}
\prod_{i<j} \exp\left[-\frac{|x_{ij}|}{a_s}\right] 
\prod_i\exp\left(-\frac{1}{2} \frac{x_i^2}{a_{ho}^2}\right)\ ,
\end{multline}
where $\mathcal{N}^{-1}$ denotes the normalization factor. As expected from Kohn's theorem, it is possible to factorize the center of mass 
$R=\frac{1}{N}\sum_i x_i$ \cite{SM},
\begin{equation}
\Psi_0(\boldsymbol{x}) =\\
\mathcal{N'}^{-1}
\exp\left(-\frac{N R^2}{2a_{ho^2}}\right) \prod_{i<j}
\exp\left[-\frac{(|x_{ij}|+Na_{ho}^2/a_s)^2}{2Na_{ho}^2}\right],
\end{equation}
where we used the identity, $\sum^N_{i} x^2_i = NR^2 + \frac{1}{N}\sum_{i<j} (x_{ij})^2$, 
and 
$\mathcal{N}'=\mathcal{N}\exp\{4a_s^2/[a_{ho}^2N^2(N-1)]\}$.

For $g>0$ (i.e. $a_s<0$) the ground state thus describes a crystal-like order in the sense that the wave function is maximal for $|x_{ij}|=Na_{ho}^2/a_s$. 

In order to analyze different physical regimes, it is convenient to recast the Hamiltonian in dimensionless form~(\ref{HLLhl})
\begin{eqnarray}
\label{Eq:H:dimensionless}
\tilde H =\sum_{i=1}^{N}\left(-\frac{1}{2}\frac{\partial^2}{\partial \tilde x_i^2}
+\frac{\tilde x_i^2}{2}\right)
+c\sum_{i<j}\left[\delta(\tilde x_{ij})
-|\tilde x_{ij}|\right]\;,
\end{eqnarray}
where tilde symbols denote that $\hbar\omega$ is used as a unit of energy and harmonic oscillator length $a_{ho}$ as a unit of distance. System properties are governed by two dimensionless parameters which are number of particles $N$ and dimensionless interaction strength $c$ defined as
\begin{equation}
c
\!=\!\frac{gm^{1/2}}{\hbar^{3/2}\omega^{1/2}}
\!=-\frac{\sigma}{\hbar^{1/2}m^{1/2}\omega^{3/2}}
\!=\!\frac{\sqrt{|g\sigma|}}{\hbar\omega}\sign(g)
\!=-\frac{2a_{ho}}{a_s}.
\end{equation}
Its value quantifies the relative strength of the interaction potential (contact and gravitational/Coulomb) with respect to the trapping potential: $c>0$ refers to repulsive contact and Coulomb potentials, while $c<0$ corresponds to attractive short- and long-range interactions.
For strong short-range attraction, $c\to-\infty$, 
the solution for the Hamiltonian~\eqref{HLL} reduces to McGuire bound-state solution 
\begin{equation}\label{Psi0HLLho}
\Psi_0(\boldsymbol{x})=\prod_{i<j}\exp\left(-\frac{|x_{ij}|}{a_s}\right)\ , 
\end{equation}
describing a quantum bright soliton. Indeed, the system energy~(\ref{Eq:E}) is similar to that of a bright soliton for $c\ll -\sqrt{24}/N$. 
In this regime, Eq. (\ref{Eq:H:dimensionless}) can be identified as the (parent) Hamiltonian with a trapped McGuire soliton as ground state \cite{delcampo20}.

{\it Mean-field theory and the Gross-Pitaevskii equation.---} In one spatial dimension, the mean-field limit is reached as the density is increased, when the distance between particles is small with respect to the scattering length. The ground state of a large number of particles is then well described by the Hartree-Fock approximation, $\Psi(x_1,\cdots,x_N) = \prod_{i=1}^N \phi(x_i)$ where $\phi(x_i)$ is a single-particle wave function. 
The variation of the free energy functional obtained for this field leads to a non-linear Schr\"{o}dinger equation~for $\Phi=\sqrt{N}\phi(x)$ \cite{Pitaevskii}
\begin{multline}\label{GPN}
\left(-\frac{\hbar^2}{2m}\frac{\partial^2}{\partial x^2}+\frac{m\omega^2}{2}x^2+g|\Phi(x)|^2\right)\Phi(x)\\
-m\alpha \int dx' |x-x'||\Phi(x')|^2\phi(x) = \mu\Phi(x)\ ,
\end{multline}
where $\mu$ is the chemical potential, $\alpha = \omega g/\hbar$ is an effective coupling constant, and the normalization condition $\int dx\ n(x) = N$ is given in terms of the local density of particle is $n(x) = |\Phi(x)|^2$. This equation can be recognized as the  Gross-Pitaevskii equation modified with an additional non-linear long-range potential 
\beqa
V(x)=-m\alpha \int dx' |x-x'||\Phi(x')|^2\ ,\nonumber
\eeqa
which is a solution of the one-dimensional Poisson equation
\beqa
\Delta V(x) = -2\alpha \rho(x)\ ,\nonumber
\eeqa
where the local mass distribution
\beqa
\rho(x)=m n(x)= m |\Phi(x)|^2\ ,\nonumber
\eeqa
 $\Delta = d^2/dx^2$ denotes the  Laplacian in $d=1$, and $\alpha$ is the coupling strength with dimensional units of acceleration. This long-range potential can be interpreted as a gravity and anti-gravity interaction between particles depending on whether the sign of $\alpha$ is negative or positive, respectively. For $\alpha>0$ the potential can also describe electrostatic interaction, after rewriting the coupling strength $|\alpha|=\hbar\omega c$, which is independent of the mass. In this case, by tuning the inverse length constant $c$, we can vary the value of the effective charge $q$ of the particle $\hbar\omega c = 4\pi \epsilon_0 r_\perp^2 |q|$, where $\epsilon_0$ is the vacuum permitivity and $r_\perp$ is a dimension reduction characteristic length.

\begin{figure}
\begin{center}
\includegraphics[width=0.49\columnwidth]{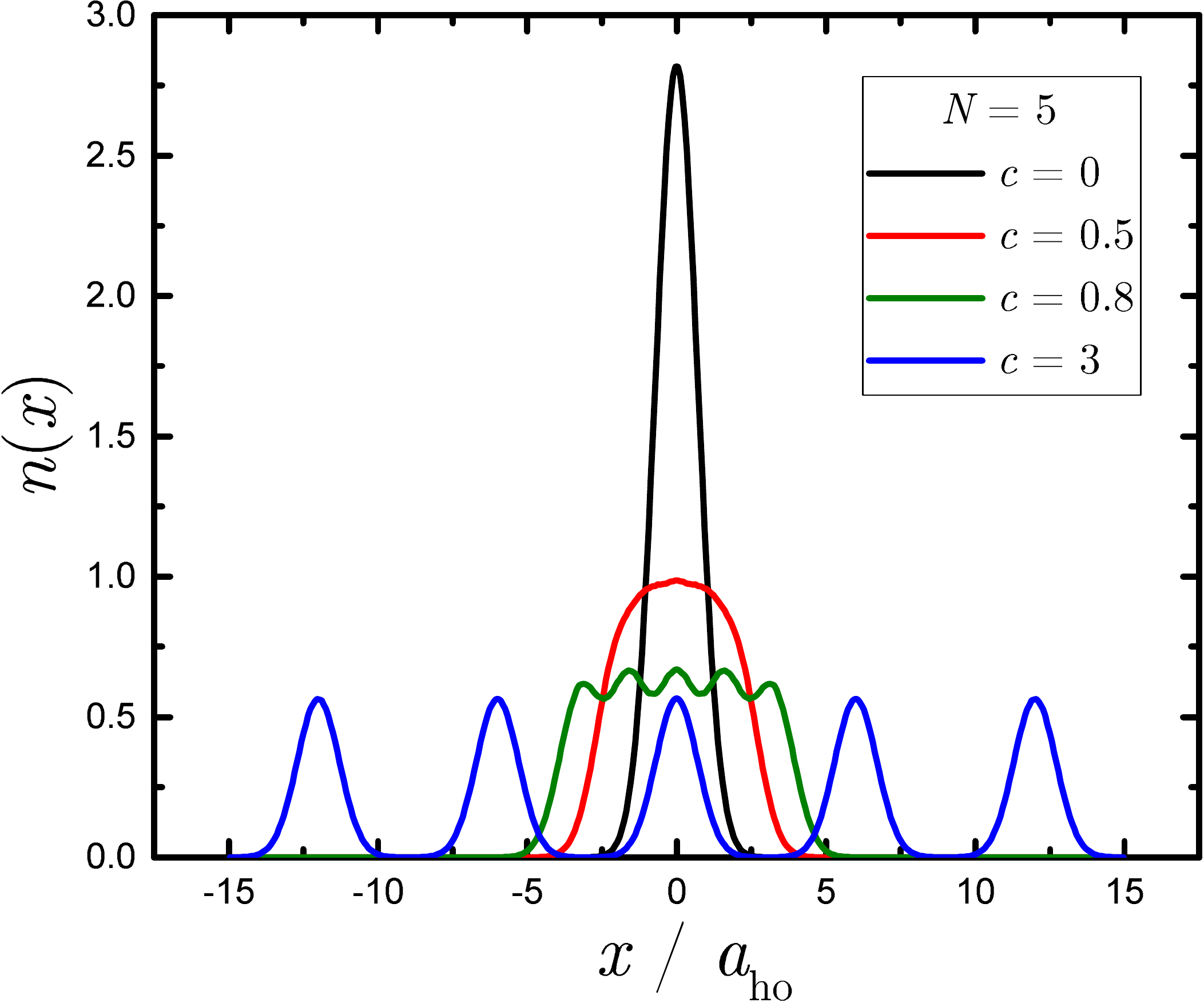}
\includegraphics[width=0.49\columnwidth]{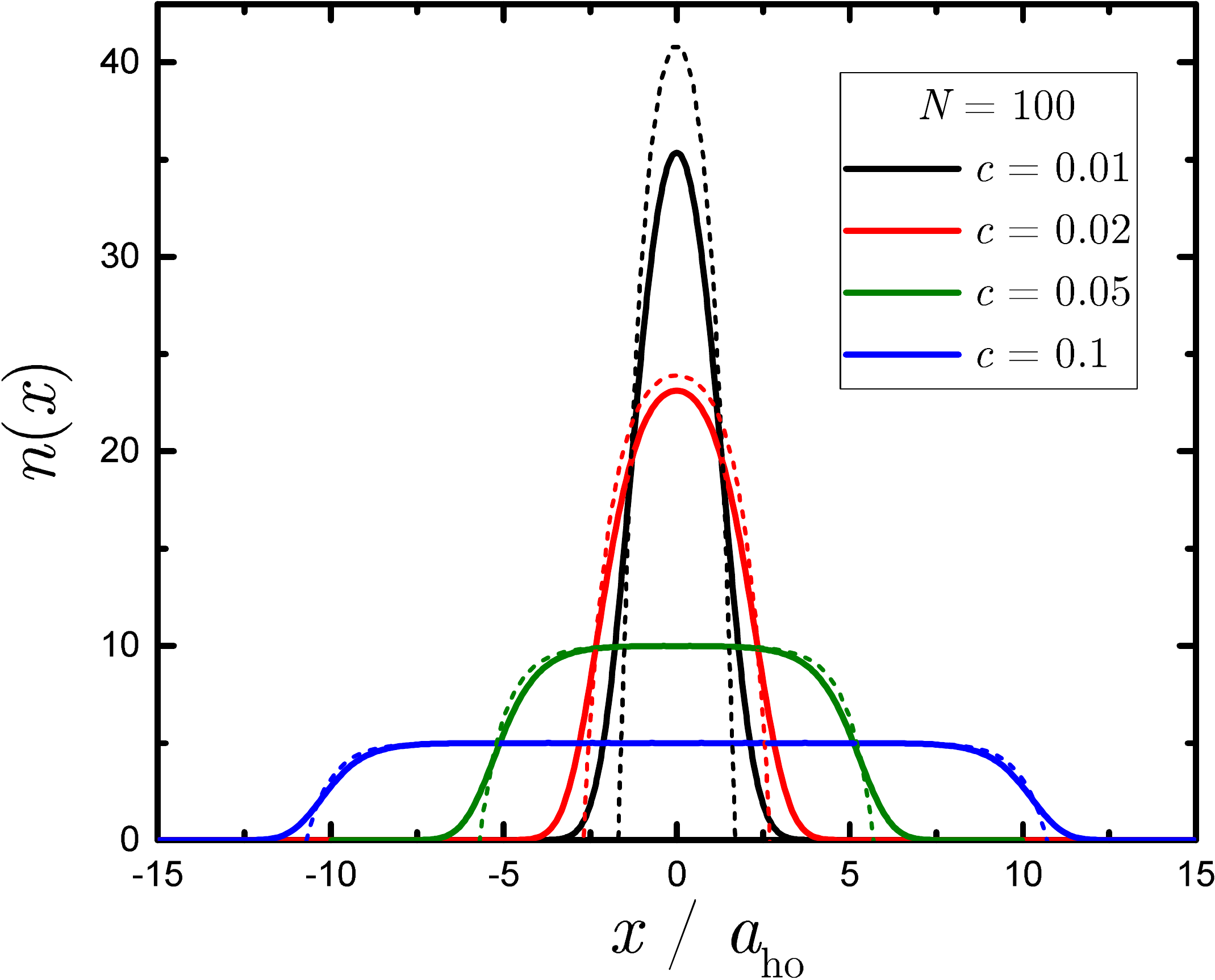}
\includegraphics[width=0.49\columnwidth]{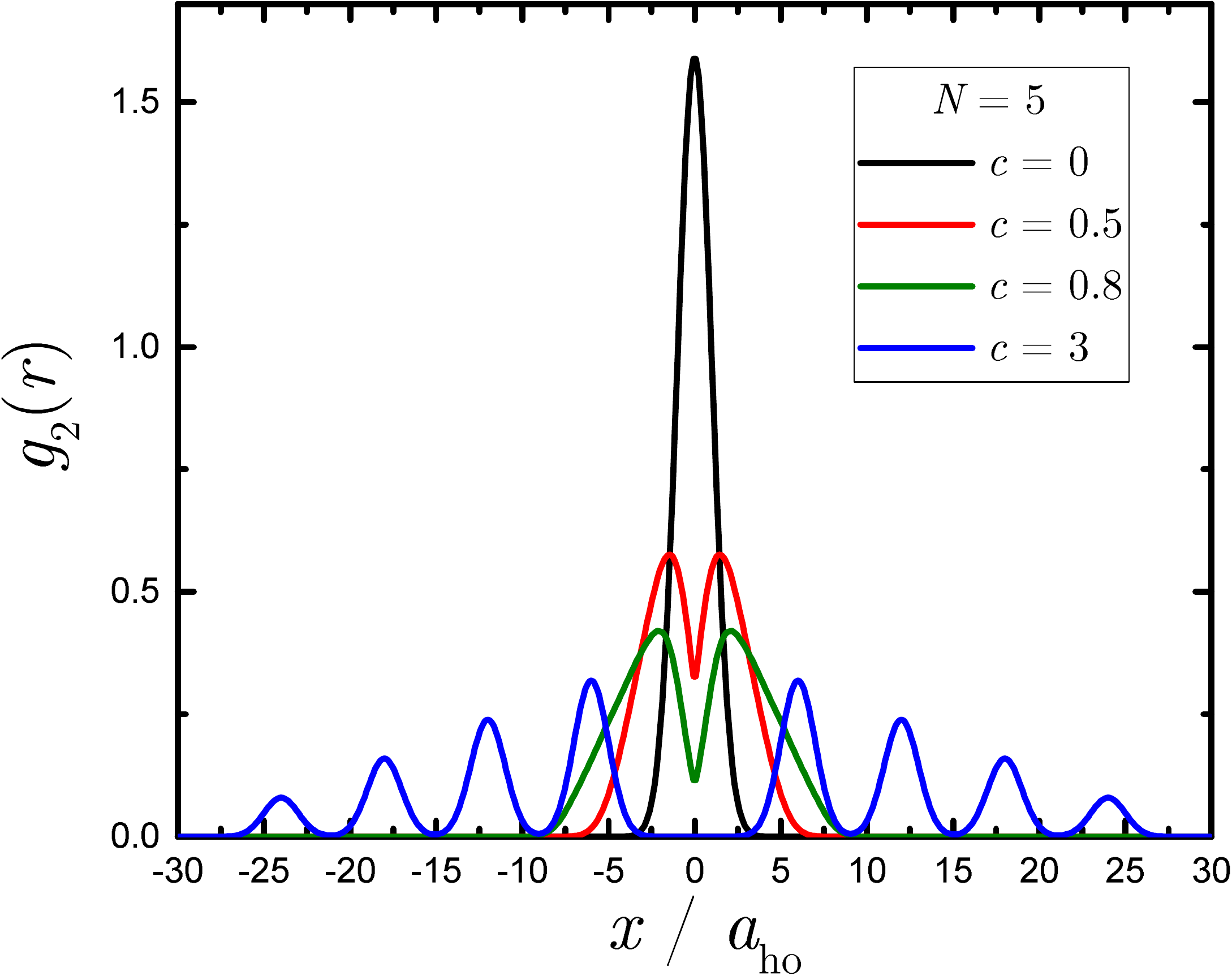}
\includegraphics[width=0.49\columnwidth]{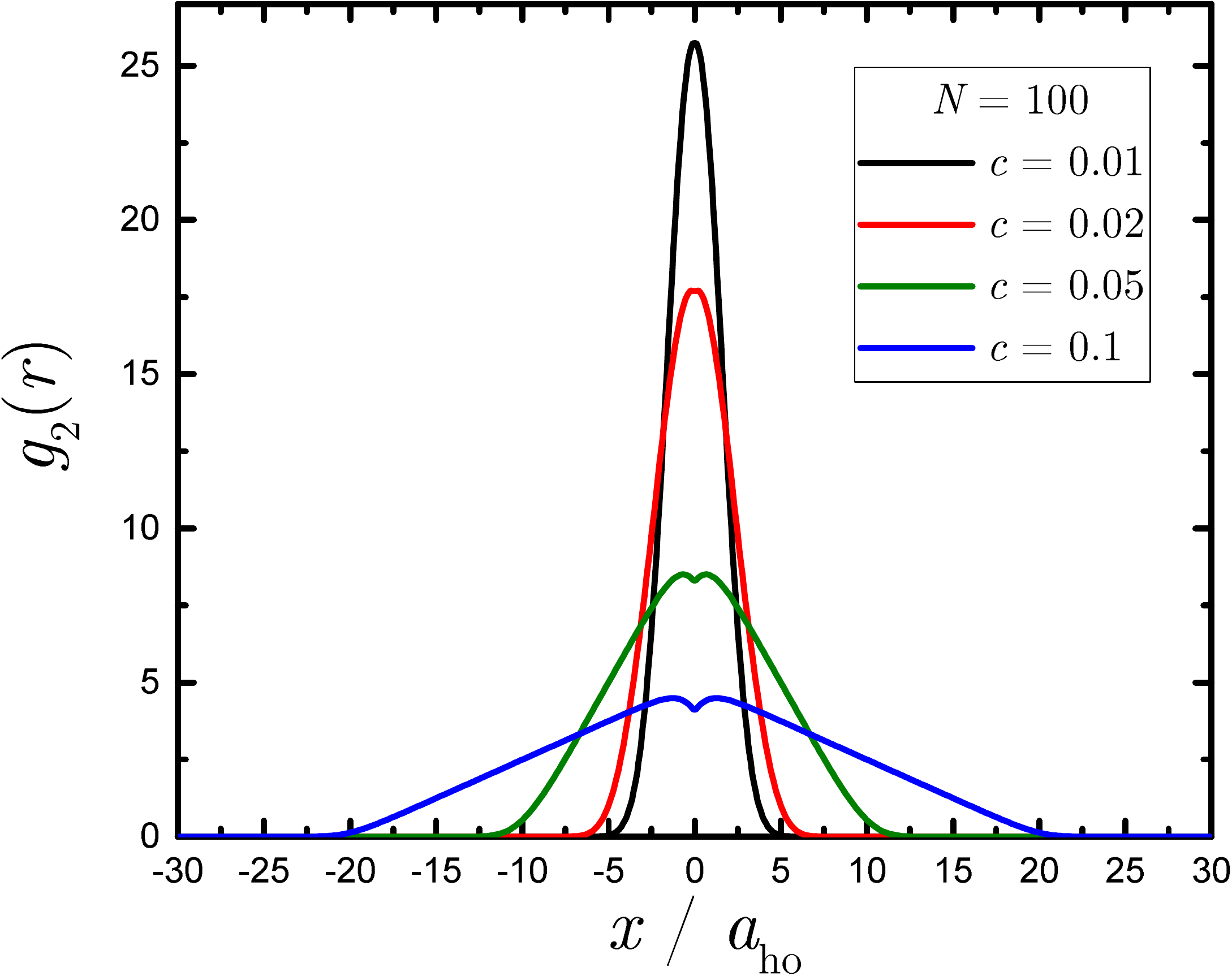}
\caption{Density profile $n(x)$ (upper row), pair distribution $g_2(x)$ (bottom row) for $N=5$ (left column) and $N=100$ (right column) particles and for different values of interaction strength.
Solid lines are results of Monte Carlo sampling, while 
dashed lines show the
mean-field profile, Eq.~(\ref{nTF}). Harmonic oscillator units are used.
}\label{Fig:Fig1}
\end{center}
\end{figure}

It follows that Eq.~\eqref{GPN} can be interpreted as a mean-field equation for a one-dimensional system of $N$ particles interacting with a short-range $\delta$- function pseudopotential and a long-range gravity ($\alpha<0$) or Coulomb/anti-gravity ($\alpha>0$) potential.
In the Thomas-Fermi regime, where the kinetic energy is negligible compared to the energy of the short-range interaction between particles, the coupling strength is greater than a critical value $g>g_c\equiv \hbar\omega a_0/(2N)$, assuming that $E_{\text{kin}}\sim\hbar^2/(2ma_0^2)$ and $E_{\text{short}}\sim g N/a_0$. In this critical regime, the energy of the long-range interaction is of the same order as that of the short-range potential, $E_{\text{long}}\sim m\omega g N a_0/\hbar=gN/a_0$, hence both short and long-range contributions are significant.  
By neglecting the kinetic term in Eq.~\eqref{GPN} we obtain the integral equation
\begin{align}
\label{nTFmu}
\nonumber n_{\text{MTF}}(x) &= \frac{1}{g}\left(\mu-\frac{m\omega^2}{2}x^2\right)&+&\frac{m\omega}{\hbar}\int dx' |x-x'|n_{\text{MTF}}(x')\\ &=\ \ \ \ \ \ \ \ \ n_{\text{TF}}(x) &+&\ \ \ \ \ \ \ \ \ \delta n(x)\ ,
\end{align} 
where $n_{\text{TF}}(x)$ is the standard Thomas-Fermi density function obtained by taking $\alpha=0$ in Eq.~\eqref{GPN}, and $\delta n(x)$ is the additional long-local term. Fortunately, Eq.~\eqref{nTFmu} can be solved using standard techniques, as shown in the Supplemental Material \cite{SM}. This yields the local density profile 
\begin{equation}\label{nTF}
n(x)a_{ho} = 
\begin{cases}
\frac{1}{2c}
\left[1-\frac{
\cosh(\sqrt{2}x/a_{ho})}
{\cosh(\sqrt{2}L/a_{ho})} \right], 
& |x|\leq L\\
  0, & |x|> L
 \end{cases},
\end{equation}
where $L$ is the size of the density profile determined by the equation $n(x)=0$ together with the normalization condition $\int_{-L}^{L}dx\; n_{\text{MTF}}(x)=N$. After numerical computation we find these values and show the density profile in Fig.~\ref{Fig:Fig1}. We also find the value of the chemical potential using Eq.~\eqref{nTFmu}. 
The normalization condition imposes that for $g\gg g_c \equiv \hbar\omega/N$ the density function is homogeneous $n_{\text{MTF}}(x)\approx\hbar\omega/(2g)$ for $|x|\leq L$ with $L\approx gN/(\hbar\omega) + a_0/\sqrt{2}$. In the opposite case with $g\ll g_c$, the cloud radius equals $L\approx \left(3g N/2m\omega^2\right)^{1/3}$, which corresponds to the standard Thomas-Fermi spread in one dimension. 

{\it Discussion.---} 
A remarkable feature of the linear long-range potential is that a negative coefficient in front of absolute value of separation $|x_{ij}|$ in Hamiltonians~(\ref{HLLhl},\ref{Eq:H:dimensionless}) actually results in repulsion rather than attraction. Indeed, upon identifying infinitely-separated particles as non-interacting, the term $-|x_{ij}|$ results in a higher potential energy at small separations.

Further, the value of the ground-state energy~(\ref{Eq:E})  is independent of the sign of $c$. This property stems from condition~(\ref{Eq:sigma}) which allow us to make the system solvable. Nonetheless, opposite signs describe drastically different regimes. For example, large value values of $|c|$ correspond to a bright soliton in the attractive (gravitational) case while they correspond to a Wigner crystal in the repulsive (Coulomb) case.

Knowledge of the exact wave function and its simple pair-product form facilitates the analysis of the ground state correlations. The local density profile $n(x)=\int dx_2\dots dx_N |\Psi_0(x_1=x,x_2,\dots x_N)|^2$ and density-density correlation function $g_2(x) = N(N-1)\int dx_1\dots dx_N \delta(x - |x_1-x_2|)|\Psi_0(x_1,\dots x_N)|^2$ can be numerically obtained by Monte Carlo integration of the square of the exact ground-state wave function. 
The obtained results are shown in Fig.~\ref{Fig:Fig1} for different characteristic values of the interaction strength $c$ and number of particles $N$. 

In the non-interacting case, $c=0$ shown with black lines in in Fig.~\ref{Fig:Fig1}, the density profile has a Gaussian shape typical for a harmonic oscillator while the pair-correlation function is continuous at the contact position, $x=0$. In the repulsive case the probability of finding two particles at the same position is reduced and a kink is formed at $x=0$ with its amplitude increasing with $c$ (compare $c=0.1$ for $N=100$ and $c=0.5;0.8$ for $N=5$ in Fig.~\ref{Fig:Fig1}). There are two types of correlations contributing to the shape of $g_2(x)$: the large $x$ envelope is dominated by the one-body density profile, while the two-body correlations provide an important contribution to the the behavior in the vicinity of $x=0$. For weak repulsion $g_2(r)$ is dominated by the one-body rather than two-body correlations and the mean-field approach is applicable. The density profile decreases monotonously from the center to the edges and for large number of particles is well predicted by the mean-field prediction of Eq.~(\ref{nTF}), shown in dashed lines. A distinguishing feature is the formation of a flat-topped mesa profile, typical of incompressible liquids where the addition of more particles does not change the density of a droplet but rather increases its size. According to Eq.~(\ref{nTF}) the density of the plateau, $n_0 a_{ho} = 1/(2c)$, is fixed by the interaction strength $c$ and the system size $L$ is directly proportional to $N$ for the large number of particles.
Yet, the system  differs from the usual liquids as the external potential is present.  The origin of its incompressibility resembles that of Laughlin states\cite{Lieb2018} as the exact solution~(\ref{Psi0HLLhl}) is analogous to the Laughlin's wave function and it allows Laughlin's plasma analogy \cite{Laughlin83}.

For strong repulsion, $c\approx 1$,  two-body correlations become important in $g_2(x)$ and result in Friedel oscillations observed in the density profile (see $c=0.5, 0.8$ and $N=5$). The mean-field approach is then no longer applicable. Finally, for even stronger repulsion, $c\gg 1$, there is vanishing probability of finding two particles in the same position, $g(0) \to 0$. The strong Coulomb interaction leads to formation of a Wigner crystal which is seen as a periodic modulation of the total density (see $c=3$ and $N=5$). Different regimes are reached via continuous crossovers in accordance to what is expected in one-dimensional and finite-size system.

In conclusion, we have introduced a model of a trapped one-dimensional Bose or Fermi atoms interacting via both contact and long-range interactions. The latter can account for either a Coulomb potential or a gravitational interaction, in the repulsive and attractive case, respectively. Our main results are the Hamiltonian~(\ref{HLLhl}), its ground state energy ~(\ref{Eq:E}) and wave function~(\ref{Psi0HLLho}), characterized by a rich phase diagram. The exact ground-state is found in a closed Laughlin-like form. For strong attractive interactions the model describes a trapped bright quantum soliton. Varying the interaction strength, the system exhibits a crossover to ideal Bose gas, incompressible-fluid mean-field regime and a Wigner crystal. The ground state density  in the mean-field regime acquires a flatted-top mesa profile which is  correctly described by a Gross-Pitaevskii equation with long-range interactions  and admits closed-form solution in the Thomas-Fermi regime. 
This rich and solvable many-particle quantum system of interacting particles under confinement should find applications in nonlinear physics, soliton theory, ultracold physics, and the description of collective quantum effects in Coulomb and gravitational systems.

{\it Acknowledgements.---} 
We acknowledge funding support from the John Templeton Foundation and UMass Boston (project P20150000029279) and the Spanish Ministerio de Ciencia e Innovaci\'on (PID2019-109007GA-I00). G.~E.~A. has been supported by the Ministerio de Economia, Industria y Competitividad (MINECO, Spain) under grant No. FIS2017-84114-C2-1-P and acknowledges financial support from Secretaria d'Universitats i Recerca del Departament d'Empresa i Coneixement de la Generalitat de Catalunya, co-funded by the European Union Regional Development Fund within the ERDF Operational Program of Catalunya (project QuantumCat, ref. 001-P-001644). 
The authors thankfully acknowledge the computer resources at Cibeles and the technical support provided by Barcelona Supercomputing Center (RES-FI-2020-2-0020).

\bibliography{MLL_lib}

\newpage

\section{Ground-state wavefunction}
Let us consider the definition of the collective coordinate
\beqa
R=\frac{1}{\sqrt{N}}\sum_{i=1}^Nx_i.
\eeqa
where we emphasized the the normalization with the $\sqrt{N}$, which will prove convenient.
We note that
\beqa
R^2&=&\frac{1}{N}\sum_{ij}x_ix_j\\
&=&
\frac{1}{N}\sum_{i}x_i^2+\frac{1}{N}\sum_{i\neq j}x_ix_j\\
&=&
\frac{1}{N}\sum_{i}x_i^2+\frac{2}{N}\sum_{i<j}x_ix_j.
\eeqa
In addition, we note that
\beqa
\frac{1}{N}\sum_{i<j}(x_{ij})^2&=&
\frac{1}{N}\sum_{i<j}(x_i^2+x_j^2-2x_ix_j)\\
&=&\frac{N-1}{N}\sum_{i}x_i^2+
\frac{2}{N}\sum_{i<j}x_ix_j.
\eeqa
In short,
\beqa
R^2=\left(1-\frac{1}{N}+\frac{1}{N}\right)\sum_{i}x_i^2-\frac{1}{N}\sum_{i<j}(x_{ij})^2
\eeqa
or simply
\beqa
\sum_{i}x_i^2=R^2+\frac{1}{N}\sum_{i<j}(x_{ij})^2.
\eeqa
As a result, one can rewrite the ground-state wavefunction as follows
\beqa
\Psi_0(\boldsymbol{x})&=&
\mathcal{N}^{-1}
\prod_{i<j} \exp\left[-\frac{|x_{ij}|}{a_s}\right] 
\prod_i\exp\left(-\frac{1}{2} \frac{x_i^2}{a_{ho}^2}\right)\\
&=&\mathcal{N}^{-1}\exp\left(-\frac{1}{2} \frac{R^2}{a_{ho}^2}\right)\prod_{i<j}
\exp\left[-\frac{|x_{ij}|}{a_s}-\frac{1}{2Na_{ho}^2}|x_{ij}|^2\right].\nonumber\\
\eeqa
Completing the square 
one finds
\beqa
\Psi_0(\boldsymbol{x})&=&
\mathcal{N}^{-1}\exp\left(-\frac{1}{2} \frac{R^2}{a_{ho}^2}\right)\prod_{i<j}
\exp\left[-\frac{1}{2Na_{ho}^2}\left(|x_{ij}|+\frac{Na_{ho}^2}{a_s}\right)^2\right]\nonumber\\
&&\times\prod_{i<j}
\exp\left[\frac{Na_{ho}^2}{2a_{s}^2}\right].
\eeqa
This yields the second expression used in the main text for the ground-state wavefunction
\beqa
\Psi_0(\boldsymbol{x})&=&
\mathcal{N'}^{-1}\exp\left(-\frac{1}{2} \frac{R^2}{a_{ho}^2}\right)\prod_{i<j}
\exp\left[-\frac{1}{2Na_{ho}^2}\left(|x_{ij}|+\frac{Na_{ho}^2}{a_s}\right)^2\right],\nonumber\\
\eeqa
with
\beqa
\mathcal{N'}^{-1}=\mathcal{N}^{-1}
\exp\left[\frac{N^2(N-1)a_{ho}^2}{4a_{s}^2}\right].
\eeqa

\section{Exact solution of Mean-field theory}

In the main body of the paper, we show that the density profile satisfies the following integral equation
\begin{equation}\label{SM:nTFmu}
\nonumber n_{\text{MTF}}(x) = \frac{1}{g}\left(\mu-\frac{m\omega^2}{2}x^2\right) + \frac{m\omega}{\hbar}\int dx' |x-x'|n_{\text{MTF}}(x')
\end{equation} 
where $n_{\text{MTF}}(x)$ is the modified Thomas-Fermi density function. In order to solve this integral equation, it suffices to take the second-order derivative on both sides of Eq. \eqref{SM:nTFmu} with respect to the variable $x$, which yields 
\begin{align*}\label{SM:nTFmu:calc}
\frac{d^2}{dx^2}\nonumber n_{\text{MTF}}(x) &= -\frac{m\omega^2}{g}+\frac{m\omega}{\hbar}\int dx'\frac{d^2}{dx^2} |x-x'|n_{\text{MTF}}(x')\\
 &= -\frac{m\omega^2}{g}+\frac{2m\omega}{\hbar}\int dx'\delta(x-x')|n_{\text{MTF}}(x')\\
 &= -\frac{m\omega^2}{g}+\frac{2m\omega}{\hbar}n_{\text{MTF}}(x)\ ,
\end{align*} 
where the commutation of the integral and  the derivative is justified by the Lesbegue theorem and we used $\frac{d^2}{dx^2}|x-x'|=2\delta(x-x')$. Therefore, we obtain a second order ordinary differential equation
\begin{equation}\label{SM:nTFmu:eq}
\frac{d^2}{dx^2}\nonumber n_{\text{MTF}}(x) = -\frac{m\omega^2}{g}+\frac{2m\omega}{\hbar}n_{\text{MTF}}(x)\ .
\end{equation} 
Using standard techniques, we find that the general solution of Eq. \eqref{SM:nTFmu:eq} is
\begin{equation}\label{SM:nTFmu:GenSol}
\nonumber n_{\text{MTF}}(x) = A \cosh\left(\sqrt{\frac{2m\omega}{\hbar}}x\right) + B \sinh\left(\sqrt{\frac{2m\omega}{\hbar}}x\right) + C\ ,
\end{equation} 
where $A,\ B,\ C$ are three real constants. The constant $C$ represents the particular solution of Eq. \eqref{SM:nTFmu:eq}, which is found to be $C = \hbar\omega/(2g)$. The constant $B$ is obviously zero as the density profile is symmetric with respect to the origin $x=0$. To find the constant $A$, we have to use the boundary condition determined by the position $x=L$ (or $x=-L$, equivalently) at which the density profile vanishes. After solving the equation $n(L)=0$ for $A$, we find 
$$A= -\frac{\hbar\omega/(2g)}{\cosh\left(\sqrt{\frac{2m\omega}{\hbar}}L\right)}\ .$$
The density function now reads  
\begin{subnumcases}{n_{\text{MTF}}(x)=\label{SM:nTF}}
\frac{\hbar\omega}{2g} \left(1-\frac{\text{Cosh}\left(\sqrt{\frac{2m\omega}{\hbar}}x\right)}{\text{Cosh}\left(\sqrt{\frac{2m\omega}{\hbar}}L\right)} \right)\ \ \ ,\ |x|\leq L \\ 
0 \ \ \ \ \ \ \ \ \ \ \ \ \ \ \ \ \ \ \ \ \ \ \ \ \ \ \ \ \ \ \ \ \ \ \ \ \ \ \ \ \ \ \ ,\ |x|>L 
\end{subnumcases}
The position $L$, which determine half of the width of the density profile, can be found using the normalization condition $\int_{-L}^{L}dx\ n_{\text{MTF}}(x)=N$, where $N$ is the number of particles. Performing the integral, we find that $L$ satisfies following equation
$$
N = \frac{\hbar\omega L}{g}\left(1-\frac{1}{\sqrt{\frac{2m\omega}{\hbar}}L}\tanh\left(\sqrt{\frac{2m\omega}{\hbar}}L\right)\right)\ ,
$$
which is a transcendental equation for $L$ that can be solved, e.g., numerically. 

\end{document}